\documentclass[preprint,nofootinbib]{revtex4}
\usepackage{amsmath,amssymb,amsbsy,latexsym}
\usepackage{hyperref}

\def\CD{{\mathcal D}}
\def\CQ{{\mathcal Q}}
\def\CR{{\mathcal R}}
\def\CS{{\mathcal S}}
\def\CT{{\mathcal T}}

\def\tr{{\rm Tr}}

\begin{document}

\title{Exact Solutions of Einstein-Yang-Mills Theory with Higher-Derivative Coupling}

\author{Hironobu Kihara}
\email{kihara(at)sci.osaka-cu.ac.jp}
\affiliation{ Osaka City University, Advanced Mathematical Institute (OCAMI),
3-3-138 Sugimoto, Sumiyoshi, Osaka 558-8585, Japan}
\author{Muneto Nitta}
\email{nitta(at)phys-h.keio.ac.jp}
\affiliation{Department of Physics, Keio University, Hiyoshi, Yokohama, Kanagawa
 223-8521, Japan
}

\preprint{OCU-PHYS 263}

\begin{abstract}
We construct a classical solution of 
an Einstein-Yang-Mills system 
with 
a fourth order term with respect to the field strength of 
the Yang-Mills field. 
The solution provides 
a compactification proposed by Cremmer and Scherk;   
ten-dimensional space-time 
with a cosmological constant
is compactified to the four-dimensional Minkowski space 
with a six-dimensional sphere $S^6$ on which 
an instanton solution exists. 
The radius of the sphere is not a modulus but 
is determined by 
the gauge coupling and the four-derivative coupling constants 
and the Newton's constant.
We also construct a solution of ten-dimensional theory 
without a cosmological constant 
compactified to AdS$_4 \times S^6$.  
\end{abstract}

\maketitle

Unification of fundamental forces with space-time and matter 
often requires higher-dimensional space-time 
rather than our four-dimensional Universe. 
The early Kaluza-Klein theory unifies 
gravity and the electro-magnetic interaction
by considering 
five-dimensional space-time with one 
direction compactified into a circle $S^1$ \cite{Nordstrom:1988fi}. 
This old idea has been revisited several times. 
After supergravity was discovered 
many people tried to unify all forces and matter 
in higher-dimensional space-time compactified 
on various internal manifolds \cite{Duff:1986hr}. 
String theory was proposed as the most attractive candidate of 
unification, but it is defined only in ten-dimensional space-time. 
In order to realize four-dimensional Universe 
one has to find a suitable six-dimensional internal space. 
So many candidates of such spaces were proposed; 
Calabi-Yau manifolds and orbifold models.
Internal manifolds can be deformed 
with satisfying the Einstein equation, 
and these degrees of freedom are called the moduli. 
The moduli introduce unwanted massless particles 
in four-dimensional world. 
Recently a new mechanism has been 
suggested to fix these moduli by 
turning on the Ramond-Ramond flux on 
the internal space \cite{Dasgupta:1999ss,Kachru:2003aw}. 
This flux compactification has been extensively 
studied in these years.

We would like to revise the compactification scenario with fixed moduli 
 proposed by
Cremmer and Scherk long time ago \cite{Cremmer:1976ir} 
(see also \cite{Horvath:1977st,Kerner:1988qj}) 
in a theory with a cosmological constant. 
By placing solitons on a compact internal space  
they showed decompactifying limit with large radius 
of the internal space is disfavored and 
the radius is fixed to a certain value 
determined by coupling constants.
They considered the 't Hooft-Polyakov monopole \cite{'tHooft:1974qc} 
on $S^2$ and
the Yang-Mills instanton \cite{Belavin:1975fg} on $S^4$, 
both of which can satisfy, with proper coupling constants, 
the first order (self-dual) equations rather than 
the second order equations of motion, 
but their solutions on higher dimensional sphere 
are not the case. 
Since string theory is defined in ten dimensions, 
it is natural to consider this scenario 
with stable BPS solitons 
on a six-dimensional internal space like $S^6$.

Higher dimensional generalization of 
self-dual equations was suggested by Tchrakian 
some years ago \cite{Tchrakian:1978sf}. 
Eight dimensional case is known as 
octonionic instantons \cite{Grossman:1984pi}.
Though several works have been done for generalized 
self-dual equations \cite{Tchrakian:1984gq,Kihara:2004yz}, 
a six-dimensional case was not discussed
because of the lack of conformal property. 
Recently we have found a new solution 
to the generalized self-dual equations
in an SO(6) pure Yang-Mills theory with 
a fourth order term with respect to the field strength of 
the Yang-Mills field (a four-derivative term) 
on a six-dimensional sphere $S^6$ \cite{Kihara:2007di}. 

In this letter we propose to use this solution 
in the context of a compactification 
of the Cremmer-Scherk type.
In our model  
ten-dimensional space-time 
with (without) a cosmological constant
is compactified to
a four-dimensional Minkowski space $M_4$ 
(anti de Sitter space AdS$_4$)
with a six-dimensional sphere $S^6$, 
where dimensionality of the internal space, six,  
is required by the four derivative term. 
Unlike the case of the absence of gravity \cite{Kihara:2007di} 
the four-derivative coupling constant $\alpha$
can differ from 
the constant $\beta$ in the generalized self-dual equations. 
When the relation $\alpha = \beta$ holds
the generalized self-dual equations 
become the Bogomol'nyi equations and solutions are BPS.
We find 
for both $M_4 \times S^6$ and AdS$_4 \times S^6$
that certain relations exist between 
the radius of $S^6$, 
the gauge coupling, the four-derivative coupling $\alpha$ 
and the gravitational coupling constants. 
When the four-derivative coupling constant $\alpha$ vanishes 
in the case of $M_4 \times S^6$,
these relations reduce to those of 
the original work 
by Cremmer and Scherk. 
The advantage of our model to the Cremmer-Scherk model is that 
the Yang-Mills soliton in our model satisfies the self-dual equations 
(the Bogomol'nyi equations for $\alpha=\beta$)
 rather than usual equations of motion 
in the case of the Cremmer-Scherk model. 
This ensures the stability of configuration 
at least for the sector of Yang-Mills fields.

Let us consider that space-time is a ten-dimensional manifold. 
We consider an Einstein-Yang-Mills theory. 
Our action contains as dynamical variables the Yang-Mills (gauge) fields $A_{\hat{\mu}}^{[ab]}$ and a graviton field or the metric $\hat{g}_{\hat{\mu}\hat{\nu}}$. Indices with a hat ``$\hat{~}$" will refer to a ten-dimensional space-time $(X^0,X^1,\cdots, X^9)$. 
Latin indices ($a,b,\cdots$) run from $1$ to $6$ and refer to an internal space. 
The Clifford algebra associated with the orthogonal group SO(6) is useful and we represent generators of the Lie algebra so(6) as their elements. The Clifford algebra is defined by gamma matrices $\{\Gamma_{a}\}$ which satisfy the following anti-commutation relations, $\{ \Gamma_a , \Gamma_b \} = 2 \delta_{ab} $. These matrices can be realized as  $8 \times 8$ matrices with complex coefficients.  
The generators of so(6) are represented by $\Gamma_{ab} = \frac{1}{2} [ \Gamma_a,\Gamma_b]$. 
We often abbreviate the Yang-Mills fields as $\displaystyle A_{\hat{\mu}}= \frac{1}{2} A_{\hat{\mu}}^{[ab]} \Gamma_{ab}$ and we also use notations with differential forms. Thus the gauge fields are expressed as $A= A_{\hat{\mu}} dX^{\hat{\mu}}$. In this notation, the corresponding field strength $F$ is written as $F = dA + e A \wedge A$, where $e$ is a gauge coupling. 
Covariant derivative $\CD_{\hat{\mu}}$ on an adjoint representation $\displaystyle Y = \frac{1}{2} Y^{[ab]} \Gamma_{ab}$ is defined as $\CD_{\hat{\mu}} Y =\partial_{\hat{\mu}} Y + e (A_{\hat{\mu}} Y  -  Y  A_{\hat{\mu}}) $, where $Y$ is a scalar multiplet. 
The action $\CS_{{\rm total}}$ consists of two parts. One is the Einstein-Hilbert action $\CS_E$ and the other $\CS_{YMT}$ is a Yang-Mills action with a term which is the fourth power of the field strength $F$. 
Such a quartic term has been studied by Tchrakian \cite{Tchrakian:1978sf} 
and so we call it the Tchrakian term. The total action is: 
\begin{align}
\CS_{{\rm total}} &= \CS_E + \CS_{YMT}~,
~~~\CS_{E} = \frac{1}{16 \pi G} \int dv   \CR  ~,\cr
 \CS_{YMT} &= \frac{1}{16} \int \tr  \left\{- F \wedge * F  + \alpha^2 (F \wedge F) \wedge *(F \wedge F) -V_0 dv \right\}~~.
\end{align}
Here the 10-form $dv$ is an invariant volume form with respect to the metric $\hat{g}$ and is written as $dv= \sqrt{-\hat{g}} d^{10}X$ in a local patch. 
The scalar curvature is denoted by $\CR$. The asterisk ``$*$" denotes the Hodge dual operator. This operator defines an inner product over differential forms, and for a given form $\omega$, $\omega \wedge * \omega$ is proportional to the invariant volume form $dv$.\footnote{The Hodge dual operator acting on a differential form on a space with Minkowski signature satisfies the following relation: $(F_{\mu \nu} dx^{\mu\nu}) \wedge *(F_{\rho \sigma} dx^{\rho\sigma})= - F_{\mu\nu}F^{\mu\nu}dv$.}  
The parameters of this action are 
the Newton's gravitational constant $G$, the gauge coupling $e$, 
the four-derivative coupling $\alpha$ and the cosmological constant $V_0$. 

 We show the explicit form of the Yang-Mills part with components of $A$ and $F$,
\begin{align}
 \CS_{YMT} &= - \int dv \left\{  \frac{1}{8} F_{\hat{\mu} \hat{\nu}}^{[ab]}F^{\hat{\mu} \hat{\nu},[ab]} + \frac{\alpha^2}{8} \tilde{T}^{[abcd]}_{\hat{\mu} \hat{\nu}\hat{\rho} \hat{\sigma}} T^{\hat{\mu} \hat{\nu}\hat{\rho} \hat{\sigma},[ab][cd]} +\frac{\alpha^2}{3\cdot 16} S_{\hat{\mu} \hat{\nu}\hat{\rho} \hat{\sigma}}S^{\hat{\mu} \hat{\nu}\hat{\rho} \hat{\sigma}} + \frac{1}{2} V_0  \right\}~~,\\
 F &= \frac{1}{4} F^{[ab]}_{\hat{\mu}\hat{\nu}} dX^{\hat{\mu}}\wedge dX^{\hat{\nu}} \Gamma_{ab}~,~~~~
S_{\hat{\mu} \hat{\nu}\hat{\rho} \hat{\sigma}} = F_{\hat{\mu} \hat{\nu}}^{[ab]} F_{\hat{\rho} \hat{\sigma}}^{[ab]} + F_{\hat{\mu} \hat{\rho}}^{[ab]} F_{\hat{\sigma} \hat{\nu}}^{[ab]} + F_{\hat{\mu} \hat{\sigma}}^{[ab]} F_{\hat{\nu} \hat{\rho}}^{[ab]}~~,
\end{align}
\begin{align}
T_{\hat{\mu} \hat{\nu}\hat{\rho} \hat{\sigma}} ^{[ab][cd]} 
&=  \left( F_{\hat{\mu} \hat{\nu}}^{[ab]} F_{\hat{\rho} \hat{\sigma}}^{[cd]} + F_{\hat{\mu} \hat{\rho}}^{[ab]} F_{\hat{\sigma} \hat{\nu}}^{[cd]} + F_{\hat{\mu} \hat{\sigma}}^{[ab]} F_{\hat{\nu} \hat{\rho}}^{[cd]}  \right) ~,\cr
\tilde{T}_{\hat{\mu} \hat{\nu}\hat{\rho} \hat{\sigma}}^{[abcd]} &= \frac{1}{6} \left( T_{\hat{\mu} \hat{\nu}\hat{\rho} \hat{\sigma}}^{[ab][cd]} + T_{\hat{\mu} \hat{\nu}\hat{\rho} \hat{\sigma}}^{[ac][db]} + T_{\hat{\mu} \hat{\nu}\hat{\rho} \hat{\sigma}}^{[ad][bc]}  + T_{\hat{\mu} \hat{\nu}\hat{\rho} \hat{\sigma}}^{[cd][ab]} + T_{\hat{\mu} \hat{\nu}\hat{\rho} \hat{\sigma}}^{[db][ac]} + T_{\hat{\mu} \hat{\nu}\hat{\rho} \hat{\sigma}}^{[bc][ad]}  \right)~.
\end{align}
The Euler-Lagrange equations from these actions 
read the usual Einstein equation and 
the equations for the Yang-Mills fields:
\begin{align}
\CR_{\hat{\mu} \hat{\nu}} - \frac{1}{2} \hat{g}_{\hat{\mu} \hat{\nu}} \CR &= {8 \pi G} \CT_{\hat{\mu} \hat{\nu}}~~,& \CD_{\hat{\mu}} \left[ \sqrt{-g} F^{\hat{\mu}\hat{\nu}} - 2 \alpha^2 \sqrt{-g}  F^{ [\hat{\mu}\hat{\nu}}  F^{ \hat{\rho}\hat{\sigma}]} F_{ \hat{\rho}\hat{\sigma}}  \right] &= 0~~.
\label{eqn:eom}
\end{align}
Here the energy-momentum tensor $\CT_{\hat{\mu} \hat{\nu}}$ is obtained by the variation of the Yang-Mills part with respect to the metric: 
\begin{align}
\CT_{\hat{\mu} \hat{\nu}} &= \frac{1}{2}  F_{\hat{\mu}}{}^{\hat{\rho},[ab]} F_{\hat{\nu} \hat{\rho}}^{[ab]} + {\alpha^2} \tilde{T}_{ \hat{\mu} \hat{\rho} \hat{\sigma}\hat{\tau} }^{[abcd]} {T}_{\hat{\nu}}{}^{ \hat{\rho} \hat{\sigma}\hat{\tau} ,[ab][cd]} + \frac{ \alpha^2}{3\cdot 2} {S}_{ \hat{\mu} \hat{\rho} \hat{\sigma}\hat{\tau} } {S}_{\hat{\nu}}{}^{ \hat{\rho} \hat{\sigma}\hat{\tau}} - \frac{1}{2} g_{\hat{\mu} \hat{\nu} } \chi \cr
\chi &=  \frac{1}{4} F_{\hat{\mu} \hat{\nu}}^{[ab]}F^{\hat{\mu} \hat{\nu},[ab]} + \frac{\alpha^2}{4} \tilde{T}^{[abcd]}_{\hat{\mu} \hat{\nu}\hat{\rho} \hat{\sigma}} T^{\hat{\mu} \hat{\nu}\hat{\rho} \hat{\sigma},[ab][cd]} + \frac{\alpha^2}{3\cdot 8} S_{\hat{\mu} \hat{\nu}\hat{\rho} \hat{\sigma}}S^{\hat{\mu} \hat{\nu}\hat{\rho} \hat{\sigma}} + V_0 ~~.
\end{align}
To solve these equations, we make an ansatz which is the same as that of Cremmer-Scherk. 
Our ansatz for the metric is the following:
\begin{align}
ds^2 &= \eta_{\mu \nu} dx^{\mu} dx^{\nu} + \frac{\delta_{IJ}}{(1+y^2/4R_0^2)^2} d y^I dy^J=\hat{g}_{\hat{\mu}\hat{\nu}} dX^{\hat{\mu}} dX^{\hat{\nu}}~~,&y^2 &= \sum_{a=1}^6 (y^I)^2~~,
\label{eqn:spheremetric} 
\end{align}
where the coordinates $X$ are the total space-time coordinates. The metric $\eta_{\mu\nu}={\rm diag}(-+++)$ is the Lorentz metric on the four-dimensional Minkowski space.  
Greek indices without a hat ``$\hat{~}$", for instance $\mu$ will refer to the first four variables. 
Capital indices ($I,J,\cdots$) run from one to six and refer to the compact space. 
The six-dimensional space is taken as a sphere with a radius $R_0$. 
The Riemann tensor, Ricci tensor and scalar curvature are
\begin{align}
R^{I}{}_{JKL} &= \frac{1}{R_0^2} \left( \delta^I_K g_{JL} - \delta^I_L g_{JK}  \right)~,~~ \CR_{IJ} = \frac{5}{R_0^2}g_{IJ}~~ ~,~~
\CR = \frac{30}{R_0^2}~~. 
\end{align}  
The rest components of the curvature tensor vanish. 
In this space, the Einstein equations in (\ref{eqn:eom}) reduce to simple equations, 
\begin{align}
- \frac{1}{2}\eta_{\mu\nu} \frac{30}{R_0^2} &= 8 \pi G \CT_{\mu\nu} ~,&0&= \CT_{\mu I}~,&
 - \frac{1}{2}\frac{ 20}{ R_0^2} g_{IJ} &= 8 \pi G \CT_{IJ} ~.
\end{align}

We now make ansatzes for the gauge fields.  
We assume that the fields $A$ do not depend on the four-dimensional directions, $\partial_{\mu} A=0$, and they have no four-dimensional components $A_{\mu}=0$. This implies that the field strengths are two forms on the six-dimensional sphere:  
$\displaystyle 
A = A_{I}(y) dy^I~, F =  \frac{1}{2} F_{IJ} dy^I \wedge dy^J
$. 
With these ansatzes, the four-dimensional part of the energy-momentum tensor becomes  $-\frac{1}{2} \eta_{\mu\nu} \chi$, and the equation reduces to 
${30}/{R_0^2} = 8 \pi G \chi$. This equation requires that the $\chi$ is a constant. 
Suppose that the field strength fulfils the 
generalized self-dual condition 
\begin{align}
F =  i \beta \gamma_7 *_6 (F \wedge F) , 
\label{eqn:self-dual}
\end{align}
where $\beta$ is a real parameter. Here ``$*_6$" means the Hodge dual on the six-dimensional sphere. 
Then the second part of the equations of motion (\ref{eqn:eom})
 is fulfilled automatically by the relation $\CD F=0$, where the exterior covariant derivative is defined as $\CD F  = d F + e \left( A \wedge F - F \wedge A \right)$. 
In fact we have an explicit solution to the self-dual equation:
\begin{align}
A &= \frac{1}{4eR_0^2} y^a e^b \Gamma_{ab}  ~~,& F &=  \frac{1}{4eR_0^2} e^a \wedge e^b \Gamma_{ab}~~,&
\beta &= \frac{eR_0^2}{3}~~.
 \label{eqn:solofgauge}
\end{align}
Here we identify the internal space index and the sphere index. 
The energy-momentum tensor of this configuration becomes
\begin{align}
\zeta & \equiv \frac{\alpha^2}{\beta^2} ~~,& \chi &= \left( 1 + \zeta \right) \frac{15}{4e^2R_0^4} + V_0~,& \CT_{IJ} &= -\frac{1}{2} \left\{ (1-\zeta) \frac{5}{4e^2 R_0^4} + {V_0} \right\} g_{IJ}~~.
\end{align}
With these ansatzes 
we obtain algebraic equations from the Einstein equations: 
\begin{align}
\frac{30}{R_0^2} &= 8 \pi G \left(  \frac{1}{2}\left( 1 + \zeta   \right) \frac{15}{2e^2R_0^4} + V_0 \right)~,&
\frac{10}{R_0^2} &= 8 \pi G \left( \left(1 - \zeta  \right) \frac{5}{8e^2R_0^4} + \frac{V_0}{2} \right)~,
\end{align}
From these we finally obtain
\begin{align}
\frac{1}{\pi G} &= \frac{1}{e^2R_0^2}  \left( 2+4 \zeta \right) 
~~,&V_0&=\frac{15}{4e^2R_0^4} \left( 1+ 3 \zeta \right)~~.
\end{align}
When the four-derivative coupling vanishes, 
$\alpha =0$ and therefore $\zeta =0$, these relations reduce to those of 
the Cremmer and Scherk \cite{Cremmer:1976ir}.
\footnote{
We need to redefine $e$ the half when we compare to the result of \cite{Cremmer:1976ir}. 
} 
When the relation $\alpha =\beta$ holds ($\zeta=1$) 
our solution saturates the Bogomol'nyi bound and becomes a BPS state. 
The energy density is given by an integral over $S^6$ as follows:
\begin{align}
E_{YMT}^{S^6} &= \frac{1}{16} \int_{S^6} \tr  \left\{- F \wedge *_6 F  + \alpha^2 (F \wedge F) \wedge *_6 (F \wedge F)  \right\}\cr
&= \frac{1}{16} \int_{S^6} {\rm Tr}  \left( i F \mp \alpha \gamma_7 *_6 (F \wedge F)
 \right) \wedge *_6 \left( i F \mp\alpha 
 \gamma_7 *_6 (F \wedge F)   \right) \pm \frac{i}{8}\alpha \int_{S^6} {\rm Tr}\gamma_7 F\wedge F\wedge
 F~~\cr
&\geq \pm \frac{i}{8}\alpha \int_{S^6} {\rm Tr}\gamma_7 F\wedge F\wedge F=  \mp \frac{1}{2^3}\int_{S^6} \epsilon_{abcdef} F^{[ab]} \wedge F^{[cd]} \wedge F^{[ef]} 
 \equiv \pm \CQ~~,
\end{align}
where the field strength $F$ has only components along $S^6$. If the coupling $\alpha$ is equal to $\beta$, the solution of eq. (\ref{eqn:self-dual}) satisfies the Bogomol'nyi equation and the energy attains the local minimum. 
We can also consider a system coupled with scalar fields. Suppose that scalar fields $Q^m$ transform as a representation of SO(6). 
The index $m$ labels the representation space. 
Let us add an action $\CS_{Q}$ of the scalar fields $Q$ 
with a Higgs potential
\begin{align}
\CS_{Q} &= \frac{1}{2} \left\{   \int dv D_{\hat{\mu}} Q^m D^{\hat{\mu}} Q^m +V(Q^2)  \right\}~~,&
D_{\hat{\mu}} Q^m &= \partial_{\hat{\mu}} Q^m  - \frac{1}{2} i e A_{\hat{\mu}}^{[ab]} R(\Gamma_{ab})_{mm'} Q^{m'}~
\end{align}
to $\CS_{{\rm total}}$.
The equations of motion are modified. In general, our solution mentioned above does not satisfy the modified equations any more. However, for the scalars which fulfil the covariantly constant condition $D_{\hat{\mu}} Q^m =0 $ and attain the absolute minimum $V(Q)=0$, the configurations of $A$ and $g$ in equations (\ref{eqn:spheremetric}), (\ref{eqn:solofgauge}) are still solutions for the modified equations. Here the constant value of the minimum is shifted to $0$. 
Thus we can argue the Higgs mechanism around our solutions.

\medskip
Next we suppose that the four-dimensional part is an anti-de Sitter 
space AdS$_4$ of radius $R_A$. 
Our ansatz for the metric is the following:
\begin{align}
ds^2 &= \eta_{\mu \nu}(x) dx^{\mu} dx^{\nu} + g_{IJ}(y) d y^I dy^J=\hat{g}_{\hat{\mu}\hat{\nu}} dX^{\hat{\mu}} dX^{\hat{\nu}}~~,
\end{align}
\begin{align}
g_{IJ}(y) dy^I dy^J &= \frac{\delta_{IJ}}{(1+y^2/4R_0^2)^2} d y^I dy^J&y^2 &= \sum_{a=1}^6 (y^I)^2~~,\cr
 \eta_{\mu \nu}(x) dx^{\mu} dx^{\nu} &= \frac{R_A^2}{\cos^2 \theta} \left( - d \tau^2 + d \theta^2 + \sin^2 \theta d \Omega^2   \right)~~,&d \Omega^2 &= \frac{|dz|^2}{(1+|z|^2/4)^2}~,
\label{eqn:adsspheremetric} 
\end{align}
where $z$ parametrizes a whole complex plane. The metric $\eta_{\mu\nu}(x)$ is a maximally symmetric metric on the four-dimensional anti-de Sitter space.  
The Riemann tensor and the Ricci tensor are
\begin{align}
R^{\mu}{}_{{\nu}{\rho}{\sigma}}&=- \frac{1}{R_A^2} \left( \delta^{\mu}_{\rho} \eta_{\nu \sigma} - \delta^{\mu}_{\sigma} \eta_{\nu \rho} \right)~,&\CR_{{\mu}{\nu}}&=-\frac{3}{R_A^2}\eta_{\mu\nu}~
,\cr
R^{I}{}_{JKL} &= \frac{1}{R_0^2} \left( \delta^I_K g_{JL} - \delta^I_L g_{JK}  \right)~,& \CR_{IJ}& = \frac{5}{R_0^2}g_{IJ}~~.
\end{align}  
The total scalar curvature is obtained by a summation of those of two parts:
$\displaystyle 
\CR = - \frac{12}{R_A^2} + \frac{30}{R_0^2}$.
In this space, the Einstein equations are
\begin{align}
\CR_{\mu\nu} - \frac{1}{2} \eta_{\mu\nu} \CR &= 8 \pi G \CT_{\mu\nu}~,& \CR_{IJ} - \frac{1}{2} g_{IJ} \CR &= 8 \pi G \CT_{IJ}~.
\end{align}
The ansatz for the gauge fields is the same as previous one and the energy momentum tensor does not change. 
With these ansatzes, we obtain algebraic equations 
from the Einstein equations as 
\begin{align}
\frac{3}{R_A^2} -  \frac{15}{R_0^2}  &= - 4 \pi G  \left\{ (1+\zeta) \frac{15}{4e^2R_0^4} +V_0 \right\}~~,&
\frac{6}{R_A^2} -  \frac{10}{R_0^2}   &=   - 4 \pi G \left\{ (1-\zeta) \frac{5}{4e^2 R_0^4} + {V_0} \right\} ~~.
\end{align}
We are interested in a possible relation to string theory and therefore 
we consider the case with the vanishing cosmological constant, $V_0=0$.  In this case, the radii ($R_A,~~R_0$) are written by the couplings,
\begin{align}
R_0^2 &= (5+7\zeta) \frac{\pi G}{4e^2}~~,&R_A^2 &= \frac{5+7\zeta}{5+15\zeta}  R_0^2 .
\end{align}
Thus the additional higher derivative coupling term of the Tchrakian type does not affect critically to the equations of motion.
When $\zeta=1$ our solution becomes a solution of the Bogomol'nyi equation again. 

Our solutions introduced in this letter are new solutions of the system with a Tchrakian term. 
The origin of this term has not been clear so far 
but it seems rather universal in order to construct 
solitons with codimensions higher than four: 
for instance it has played a crucial role 
to construct a finite energy monopole (with codimension five) 
in a six-dimensional space-time \cite{Kihara:2004yz}.
Though the parameter $\zeta (=\alpha^2/\beta^2)$ is a free parameter, 
we expect that the system goes to $\zeta=1$ because it becomes BPS.
There are several discussions on the (in)stability of higher-dimensional Yang-Mills theories \cite{Randjbar-Daemi:1983bw}. 
To compute the mass spectra of the fluctuations around our solutions is a future work. 
When the scalar fields $Q^m$ are non-trivially coupled, 
the system may allow  BPS composite solitons 
which are made of solitons with different codimensions, 
as in the case of usual self-dual Yang-Mills equations 
coupled to Higgs fields 
\cite{Eto:2006pg}.

Finally our solution of AdS$_4 \times S^6$ may have a relation with 
D2-branes, and we hope that there exists some impact on AdS/CFT duality 
\cite{Maldacena:1997re}.

\begin{acknowledgments}
We are grateful to D.~H.~ Tchrakian for various comments.  
We would like to thank Y. Hosotani, H. Itoyama, Y. Yasui, M. Sakaguchi, T. Oota, T. Kimura, S. Shimasaki and E. Itou.  
We also thank M. Sheikh-Jabbari for an advice. 
This work is supported by the 21 COE program ``Constitution of wide-angle mathematical basis focused on knots" from Japan Ministry of Education. 
\end{acknowledgments}

\end{document}